**Title:**

Evidence for 2D Solitary Sound Waves in a Lipid Controlled Interface and

its Biological Implications

**Authors:** Shamit Shrivastava[1], Matthias F. Schneider[2]*

**Affiliations:**

[1]Department of Biomedical Engineering, Boston University, Boston, MA 02215, USA.

[2]Department of Mechanical Engineering, Boston University, Boston, MA 02215, USA.

*Correspondence to: mfs@bu.edu

**Abstract**: Biological membranes by virtue of their elastic properties should be capable of propagating localized perturbations analogous to sound waves. However, the existence and the possible role of such waves in communication in biology remains unexplored. Here we report the first observations of 2D solitary elastic pulses in lipid interfaces, excited mechanically and detected by FRET. We demonstrate that the nonlinearity near a maximum in the susceptibility of the lipid monolayer results in solitary pulses that also have a threshold for excitation. These experiments clearly demonstrate that the state of the interface regulates the propagation of pulses both qualitatively and quantitatively. We elaborate on the striking similarity of the observed phenomenon to nerve pulse propagation and a thermodynamic basis of cell signaling in general.

**Introduction:** In a living system, hydration shells around membranes, biomacromolecules (*e.g.* proteins, DNA, etc.) and ions form quasi-2D interfacial zones that account for most of the interstitial water. Such an interface can be characterized thermodynamically by its state diagrams [1] that map the interrelationships of physical variables (lateral pressure ↔ surface density, surface potential ↔ charge, etc.). These state diagrams can be obtained *experimentally* from macroscopic measurements on the interface. One of us has previously established that the state of the interface is a crucial determinant for equilibrium and non-equilibrium events at the interface [2–4]. For instance, heat capacity or compressibility of the interface[2–4] has been shown to regulate trans-membrane current fluctuations or 2D pulse propagation, during which all the variables of interface (pressure, temperature, surface-potential, fluorescence, density, charge etc.) are affected simultaneously as required by Maxwell's relations [3,5,6]. Such events derived from fundamental physical principles may imply certain biological functions (e.g. local membrane transport and communication) and this relationship between state and function can be tested experimentally. For example, sound waves at the membrane interface have been proposed as the physical basis of nerve pulses and it is believed that key features of action potentials (shape stability, all-or-none nature, etc.) result from nonlinear properties of cell membranes [7–10], an idea first put forward by K. Kaufmann[10]. Theoretically it's not the complexity of membrane composition, which includes proteins (ion channels and pumps), lipid heterogeneity etc., but the nonlinearity in the elasticity of the interface that is necessary to support such sound waves. Although such conditions on elasticity are met even in a single or multiple component lipid system, the existence of nonlinear sound waves, also known as solitary elastic waves, that resemble action potentials, has not been demonstrated in a hydrated lipid interface, yet.

This study reports the first observation of "solitary" elastic waves in lipid monolayers. Propagating waves were excited mechanically and the resulting propagating variations in the state were detected by FRET measurements. The relationship between the thermodynamic state and velocities of the propagating waves confirmed that they indeed travel within the lipid interface [4]. Most importantly, only when the system was close to the nonlinear regime of the state diagram, sharp localized pulses with large amplitudes were observed. Furthermore these pulses were excitable only above a critical state-dependent threshold. These self-supporting solitary wave packets propagating in the interface bear striking resemblances to action potentials in living systems.

**Opto-Mechanical Experiments on a Lipid Interface**

From an experimental point of view, the ability to control and monitor the state of an interface is crucial for a thermodynamic approach. This is easily achieved in lipid monolayers where the diagrams of state, for instance lateral pressure or surface potential vs. area, can be obtained at various bulk pH and temperature conditions [11,12]. We recently showed that the fluorescence intensity of dye molecules embedded in the lipid monolayer is also a thermodynamic observable of the interface [6], meaning that it can be analyzed as any other thermodynamic property of the interface (such as surface potential, density etc) once the equation of state in terms of fluorescence emission is known. Fluorescence measurements provide a fast and noncontact means for measuring state changes associated with propagating pulses. In this study we use FRET between a pair of dye molecules, which has two major advantages over standard fluorescence intensity measurements; (i) ratiometric measurements provide significantly better signal to noise ratios [13] and (ii) one can distinguish between longitudinal (compressional) and transversal (capillary) components of a pulse [14] (Fig. 1 and S1 Comment). The experimental setup shown in Fig. 1 was used to control and monitor the state of the lipid monolayer at the air/water interface. A chloroform solution, containing lipids Dipalmitoylphosphatidylcholine (DPPC), the donor *N*-(7-Nitrobenz-2-Oxa-1,3-Diazol-4-yl)-1,2-Dihexadecanoyl-*sn*-Glycero-3-Phosphoethanolamine, Triethylammonium Salt (NBD-PE) and acceptor dye molecules Texas Red® 1,2-Dihexadecanoyl-*sn*-Glycero-3-Phosphoethanolamine, Triethylammonium Salt (Texas Red® DHPE) from Invitrogen (100:1:1) was spread on the water surface of a Langmuir trough. The mean lateral pressure ($\boldsymbol{\pi}$) of the monolayer was measured using a Wilhelmy plate. FRET between donor and acceptor dye molecules embedded in the lipid monolayer was measured ratiometrically (20kS/s) by the FRET parameter defined as;

$$\theta = \left(\frac{I_{535nm}}{I_{605nm}}\right) \tag{1}$$

For this, emission intensities were acquired simultaneously at two wavelengths (535 and 605nm) while fluorescence was excited at 465nm. A razor blade attached to a piezo cantilever was arranged such that the long edge of the blade touches the air water interface forming a meniscus, while the motion of the cantilever moves the blade horizontally along the interface. The pulses were excited by using controlled mechanical impulses ($\Delta t \sim 10ms$) of the piezo cantilever (Fig. 1). The maximum displacement (~1mm) of the blade represents the upper bound for the amplitude on a relative scale of 0-1A which can be tuned using the power supply for the piezo element. This technique has previously been shown to ensure maximum coupling to the longitudinal mode [15]. The ensuing changes in state were quantified optically via the relative changes in FRET parameter derived from eq.1 (Fig 1);

$$\frac{\Delta\theta}{\theta} = \frac{\Delta I_{535}}{I_{535}} - \frac{\Delta I_{605}}{I_{605}} \tag{2}$$

In order to understand the state dependence, the pulses were excited at different mean surface pressures.

**Localized Nonlinear Pulses Controlled by State of the Interface**

Figure 2 presents the first observations of nonlinear pulses propagating along a lipid monolayer. The pulses were measured optically using Eq.2 (also see Fig. S1) in fundamentally different regimes of the state diagram (marked in Fig. 2D and 2E). The regimes differ in terms of their isothermal compressibility $\kappa_T = -\frac{1}{A}\left(\frac{\partial A}{\partial \pi}\right)_T$ that can be obtained from the $\pi \leftrightarrow A$ isotherm (inset Fig. 2E). $\kappa_T$ has a sharp peak in the transition region around $\pi$=6mN/m marked as regime (ii), while regime (i) at $\pi$=3.2mN/m and regime (iii) at $\pi$=12.8mN/m represent the two states that lie far on either side of the peak in $\kappa_T$ . Comparing the pulse shapes in these three different states reveals *at least* two striking differences: (i) for the same strength of excitation the amplitude increases by 1 to 2 orders of magnitude (~40 times) when excited in the transition regime and (ii) a significant temporal confinement or steepening of the pulse appears simultaneously (Fig. 2A, 2B and 2C). The latter can be better appreciated in the frequency domain representation where the temporal confinement inversely results in a significant spectral broadening ($\Delta\omega \sim 1/\Delta t$), as will be discussed below. The exact pulse shape varies strongly as a function of state near the transition and can be resolved further (Fig. S2).

**Threshold Amplitude and "All-or-None" Nature**

The sensitive dependence of pulse's shape on state combined with the abrupt increase in its amplitude and confinement indicates that the observed phenomenon is of nonlinear nature. To analyze this nonlinearity further pulse amplitude $\Delta\theta$ and shape (frequency spectrum) were obtained as a function of excitation strength (peizo amplitude). This experiment presented another remarkable feature of the observed solitary pulses, the presence of a threshold. Only excitations of amplitude above a certain threshold could excite solitary pulses (Fig. 3). On zooming in to the pulse shape, it is observed that near the threshold, the increased excitation strength is temporally focused onto certain regions along the pulse shape (Fig. S4). This results in excitation dependent steepening and confinement of the pulse's shape along with a nonlinear increase in its amplitude. This is better represented by the broadening of the corresponding frequency spectrums (FFTs) (Fig. 3C). In a linear system, stronger excitations simply cause elevations or rescaling of spectral features to higher amplitudes [16]. However, clear qualitative changes in the spectrum are observed for the pulses presented herein, which underline the impact of nonlinear effects. The higher frequencies (100-300Hz), present at an excitation of 0.83A, are practically absent at 0.67A, where A is an arbitrary scale between 0-1 representing the strength of excitation. However, a similar rise in excitation strength from 0.50A to 0.67A had practically no effect, as represented by near complete overlap of the corresponding spectrums. The onset of higher frequency components near 0.77A (also see Fig. S3) and subsequent strong spectral broadening indicate the existence of a threshold. Although the amplitude dependent spectral broadening observed here is typical of systems with nonlinear susceptibilities [16,17], the precise control over the degree of nonlinearity is an extraordinary advantage of the present setup.

**Theoretical Considerations**

Theoretically, the nonlinearity in the adiabatic compressibility $\kappa_s$ of the interface relates to the evolution of a propagating pulse at the interface [4,6]. $\kappa_s$ is related to the velocity of propagation $c_g$, which has been measured ($c_{exp}$) and plotted as a function of mean pressure and mean FRET parameter $\theta_0$ (Fig. 4). The minimum in $c_{exp}$ is related to a maximum in $\kappa_s$ (see Fig. 2E and Fig. 4) [4,6] confirming the compressional nature of the optically observed pulses. The propagation velocity of a wave packet is represented by the group velocity $c_g(\theta, \omega) \equiv \frac{\partial \omega}{\partial k}$ which in general is a function of both the amplitude $\theta$ and frequency $\omega$, k being the wave number. The time delay $\Delta t$, measured experimentally, is inversely related to $c_g$ and in a dispersive nonlinear system $\Delta t$ varries within a single pulse $\Delta\theta(t)$ that usually leads to broadening of the pulse shape as it travels. For a pulse of amplitude $\Delta\theta$ and spectral width $\Delta\omega$, the broadening scales as

$$\Delta\Delta t \sim \Delta \frac{1}{c_g} \sim \frac{\partial}{\partial \omega} \frac{1}{c_g} \Delta\omega + \frac{\partial}{\partial \theta} \frac{1}{c_g} \Delta\theta \sim \beta\Delta\omega - \frac{c_g'}{c_g^2}\Delta\theta \qquad (3)$$

, where $\beta \equiv \frac{\partial^2 k}{\partial \omega^2}$ is known as the group velocity dispersion (GVD) parameter [18] and $c_g' = \frac{\partial c_g}{\partial \theta}$ represents the nonlinearity in compressibility. Qualitatively $c_g'$ can be estimated from the phenomenological dependence $c_g(\theta) \sim c_{exp}(\theta)$ (Fig.4), where the tangent on any point along the curve would be directly related to $c_g'$. For a preserved pulse shape ($\Delta\Delta t \rightarrow 0$) a simple relation, representing the balance between nonlinearity and dispersion can be derived as $\beta\Delta\omega \sim \frac{c_g'}{c_g^2}\Delta\theta$. This relationship can be compared with experiments to test if the observed nonlinear pulses are indeed of solitary nature. In Fig 3A it's reasonable to assume that the values for $\beta$ and $\frac{c_g'}{c_g^2}$ do not change significantly between the two super-threshold pulses. On comparing the spectral half widths (Fig. 3C) and amplitudes (Fig. 3A) of these two pulses $\left|\frac{\Delta\theta}{\Delta\omega}\right|_{0.83} \sim \left|\frac{\Delta\theta}{\Delta\omega}\right|_{1.00} = 0.0045$ indicating that the pulse shape is indeed conserved as the nonlinearity and dispersion balance each other. In addition to the solitary nature of the observed pulses, the relation between dispersion and nonlinearity also explains the origin of the threshold. The slope $c_g'$ of the phenomenological curve $c_g(\theta)$ (Fig.4) changes sign (negative to positive) on increasing pressure across the minimum in $c_g(\theta)$. Assuming $\beta$ doesn't change its sign, the dispersion and the nonlinearity can only be balanced on one side of the minimum, while on the other side the two effects would rather reinforce each other, smearing out any pulse propagation immediately (note that $\Delta\theta$ is observed to be positive in this region). This is further supported by resolving the state dependence of the pulse shape more sensitively (Fig. S2). Upon a small increment in $\pi(A)$ from 5.1mN/m to $\pi(A) = 5.3$mN/m the signal shoots up nonlinearly indicating a strong threshold when approaching the transition from the liquid expanded side. In conclusion not only the observed pulses are of solitary nature, they also have a clear threshold for excitation due to the second order nonlinear effects at the minimum in $c_g$ near the transition.

Further experiments in different lipid systems are required for a deeper understanding of the observed nonlinearity and threshold. Nonlinearity, dispersion and viscosity, which all depend on state [19,20], affect the evolution of pulse (shape) and hence need to be systematically examined. As discussed for a solitary pulse of given amplitude the width is conserved despite dispersion when nonlinear effects of the medium counteract sufficiently. However in a dissipating medium the decay in amplitude will result in a corresponding broadening of the pulse to the point where the nonlinearity cannot balance the dispersion anymore, i.e. that during propagation the amplitude will eventually "slip" below the threshold due to dissipation. This is clearly observed for the solitary pulses reported in this study (Fig. S5), although at this point the primary reason for decay in amplitude is not clear (we imagine dissipative or geometrical effects as a possible source) (see Fig.1). Furthermore, the role of spatial confinements [21] and the behavior of two pulses under collision will also be explored. In nonlinear systems, two pulses might change or annihilate on collision (like action potentials) or remain unaffected like solitons [7,22–24].

**Biological Implications**

So far we have tried to emphasize that the solitary pulses with a threshold for excitation result from a peak in compressibility and we have shown this systematically for a lipid monolayer. However, the same line of arguments applies for lipid bilayers and real biological membranes as long as their equation(s) of state give rise to nonlinearities [11,25–28]. The later is experimentally well established in living cells and single neurons [25,28] and as the measurement of the susceptibilities ($c_p, \kappa_T$ etc.) of living systems (single cells) improve, the relevance of such nonlinear phenomena in biology will be better established. The obvious differences between mono and bilayer – although highly important for a detailed analysis – are therefore irrelevant from a fundamental point of view. Nevertheless, a brief discussion on mono v/s bilayer as acoustic mediums may be helpful. The problem of mono and bilayer correspondence is steeped in rich literature [29,30] but its most important aspect relevant to sound propagation is probably the role of surface tension which is significantly high ($\gamma \sim 30 \leftrightarrow 72mN/m$) in monolayers at air/water compared to a leaflet in a bilayer where $\gamma \rightarrow 0$. This has direct implications for capillary (transversal) modes which scale with $\gamma$ and are therefore almost unavoidable in monolayers but are likely to be absent in a bilayer [29,30]. Excitation of longitudinal waves is therefore expected to be more efficient in lipid bilayers. Still, the crucial difference in boundary conditions in the two systems and the role of chemical gradient across the membrane and its integration into a propagating state change need to be further investigated [3].

Finally based on the general role of nonlinear state diagrams in our study and given the recent discussion on the thermodynamic foundation of nerve pulses, we would like to discuss the biological relevance of our findings. The shape of the solitary pulses in lipid monolayers and action potentials in cell membranes can be directly compared because fluorescence reports membrane potential in both cases [6,31,32]. Based on this correspondence, the solitary waves excited in the transition regime with a relative FRET amplitude $\Delta\theta/\theta$ of 1.2 units are estimated to measure ~200mV in surface potential while sub-threshold pulses measured outside this region will correspond to a surface potential of approximately 5mV [5]. There are several striking similarities between our results on lipid monolayers and the data on nerve pulses: (i) both systems support 'all-or-none' pulses which propagate as solitary waves and exist only in a narrow window bound by certain nonlinearities in their respective state diagrams[28,33,34], (ii) the pulses in both systems represent an adiabatic phenomenon [35,36] and are not only electrical but are inseparably mechanical (deflection and volume), optical (polarization, chirality, fluorescence, turbidity) and thermal(temperature, enthalpy) pulses also [5,6,32,33,35,37–42]

The velocities of the pulses reported herein are in the same order of magnitude as those reported for action potentials in plant cells as well as non-myelinated animal cells [43–45]. Similarly the biphasic shape (see Fig. 2B) is quite characteristic of action potentials and its similarity to the observed pulses in this study is striking and should be compared with Fig. 13 in Hodgkin and Huxley's famous work[46]. Interestingly the biphasic pulse shape obtained in the monolayer does not require separate proteins and accompanying ion fluxes to explain different phases (rising, falling, undershoot) as in an action potential. Despite these similarities, however, we believe that absolute velocity and pulse shape are not proper criteria for testing a new theory of nerve pulse propagation as both vary tremendously, in cells and as well as in lipid monolayers, depending on composition, excitation and state of the membrane interface [47–49]. Rather it's the variation in velocity as a function of state $c_g(\pi, T)$, variation in pulse shape as a function of degree of nonlinearity $c_g'(\pi, T)$ and the existence of a threshold that can be explained thermodynamically, as seen in this study.

Finally, given that the state of the interface has been shown to strongly correlate with the activity of membrane-bound proteins and enzymes [50–52], we are also looking at the effect of these pulses on protein and enzyme kinetics as a new mechanism in biological signaling. Further studies will show whether the solitary elastic pulses as reported here are indeed a physical basis of nerve pulses and cellular communication in general.

**Acknowledgments:** MFS thanks Dr. Konrad Kauffman (Göttingen) who first introduced him to the thermodynamic origin of nerve pulse propagation and its theoretical explanation. We would also like to thank him for numerous seminars and discussions. We would like to thank Dr. Christian Fillafer and Dr. Josef Griesbauer for helpful discussions. Financial support by BU-ENG-ME and BU-XTNC is acknowledged. MFS appreciates funds for guest professorship from the German research foundation (DFG), SHENC-research unit FOR 1543.

## Figures

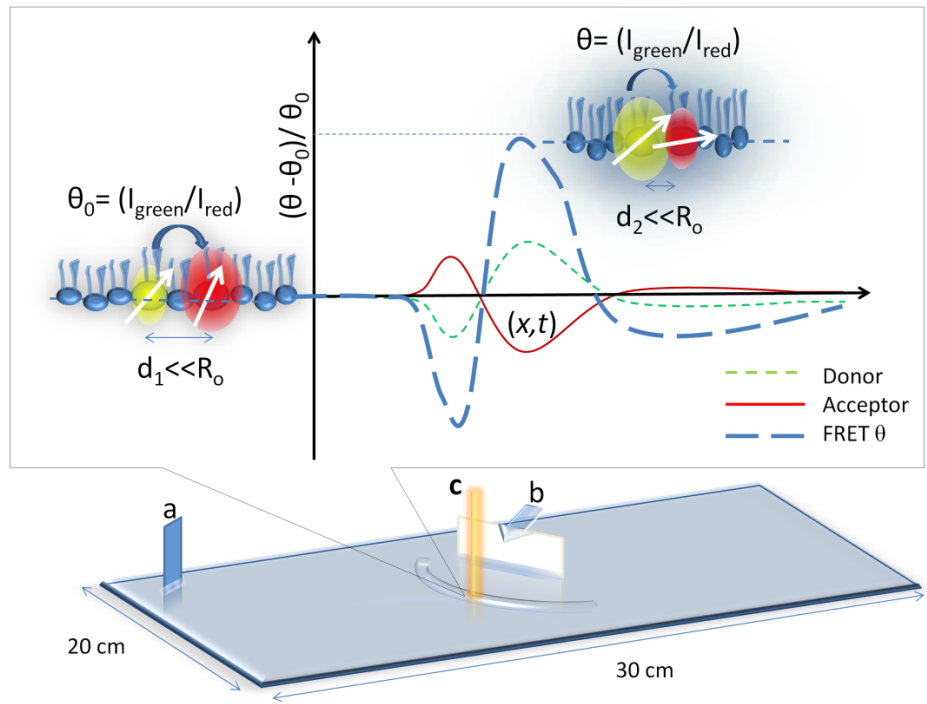

**Figure 1. Experimental setup.** The state of the interface was controlled by titrating a lipid-dye solution on a Langmuir trough while measuring the lateral pressure (at a). A razor blade (b) was actuated horizontally by a piezoelectric element in order to excite longitudinal pulses that were detected by FRET at c, where distance bc=1cm. The cartoon represents a microscopic interpretation of FRET at an interface. In an ordered 2D medium, FRET efficiency $\sim f(r, \overrightarrow{\mu_1} . \overrightarrow{\mu_2})$, where r is the distance and μ represent the emission or adsorption dipole respectively. Therefore along with distance, the relative orientation of absorption and emission dipoles is also relevant as opposed to their absolute orientation with respect to the interface. As the pulse arrive the relative distance and orientation of the transition dipoles of the donor and acceptor molecule change as a function of state represented by $(\theta, \pi)$ [6]. Correspondingly the FRET efficiency changes which leads to anticorrelated changes in the intensity of donor (green small dash) signal and acceptor (red solid) signal. On the other hand any motion of the interface due to capillary modes of the water waves travelling simultaneously with the longitudinal modes will change the donor and acceptor signal in a correlated manner. The FRET parameter $\frac{\Delta\theta}{\theta_o}$ (blue, long dash) amplifies the anticorrelated parts of the signal while filtering out the correlated parts. Finally it is the relationship of the observed velocities to the compressibility of the interface that confirms compressional nature of the waves (see Fig. 4).

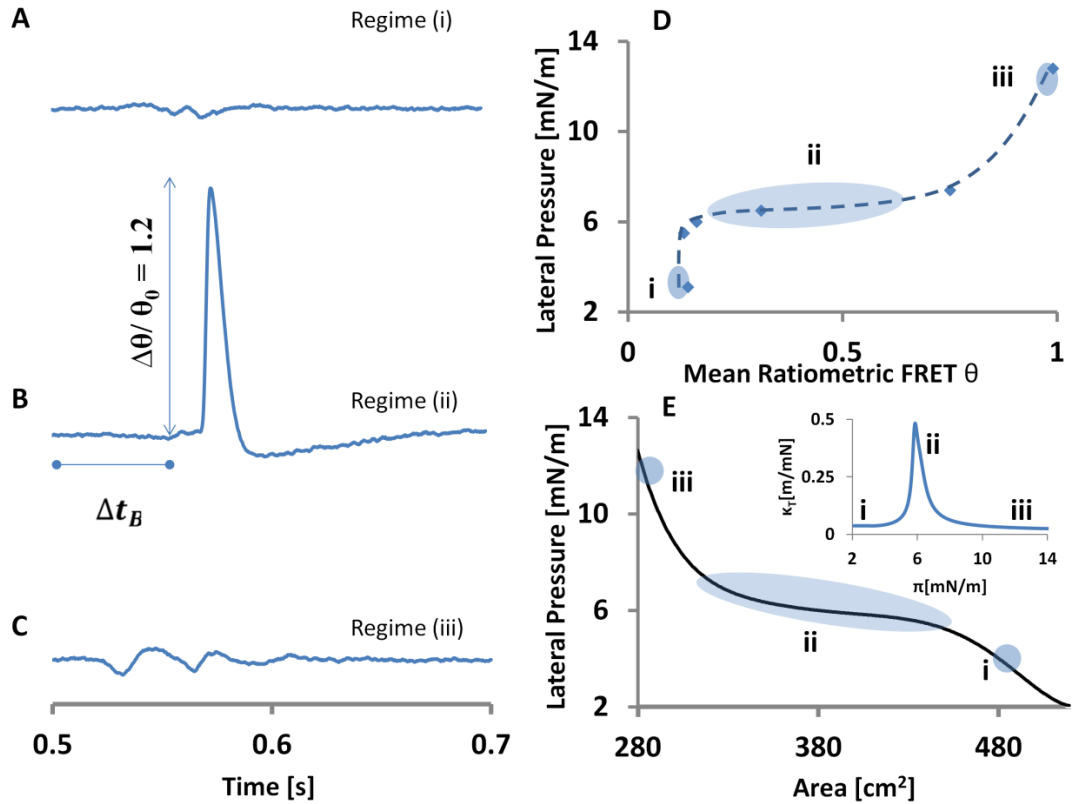

**Figure 2. State dependence of pulse shapes in lipid interface.** Elastic interfacial pulses **(A ,B ,C)** were excited in fundamentally different regimes (i, ii and iii) of the state diagram **(D,E)**[6] at $\pi(A)$=3.2 mN/m, 6mN/m and 12.8mN/m respectively. The inset in (E) shows the three regimes on the isothermal compressibility $\kappa_T(\pi)$ curve calculated from $\pi(A)$. A distinct pulse **(B)** only appears near or in the plateau (regime ii) of the state diagram and is an order of magnitude stronger in amplitude (~40 times) than pulses in regimes (i) and (ii) (see Fig. S1 for individual donor and acceptor signals). Although the sharp pulse as in **(B)** only appears near the nonlinear regime of the state diagram, the absolute shape is very sensitive to the precise state along (**i**) to (**ii**) (See Fig. S2). In these experiments the state was altered by changing the surface density of lipid molecules but it can equally well be varied by changing other physical parameters (temperature, pH, lipid-type, ion or protein adsorption, solvent incorporation etc.). The delay $\Delta t$ was used to calculate the experimental propagation velocity $c_{exp}$. Experimental details: pulses measured via FRET (see eq.2 and Fig. S1), lipid (DPPC) monolayer at 19 $^o$C for the excitation resulting from a piezo amplitude of ~ 1mm (1.5A), distance between excitation and detection was 1cm. The dashed line in (D) is a guide to the eye. The similarity of the pulses in the nonlinear regime **(ii)** with action potentials (see figure 13 in [46]) is striking.

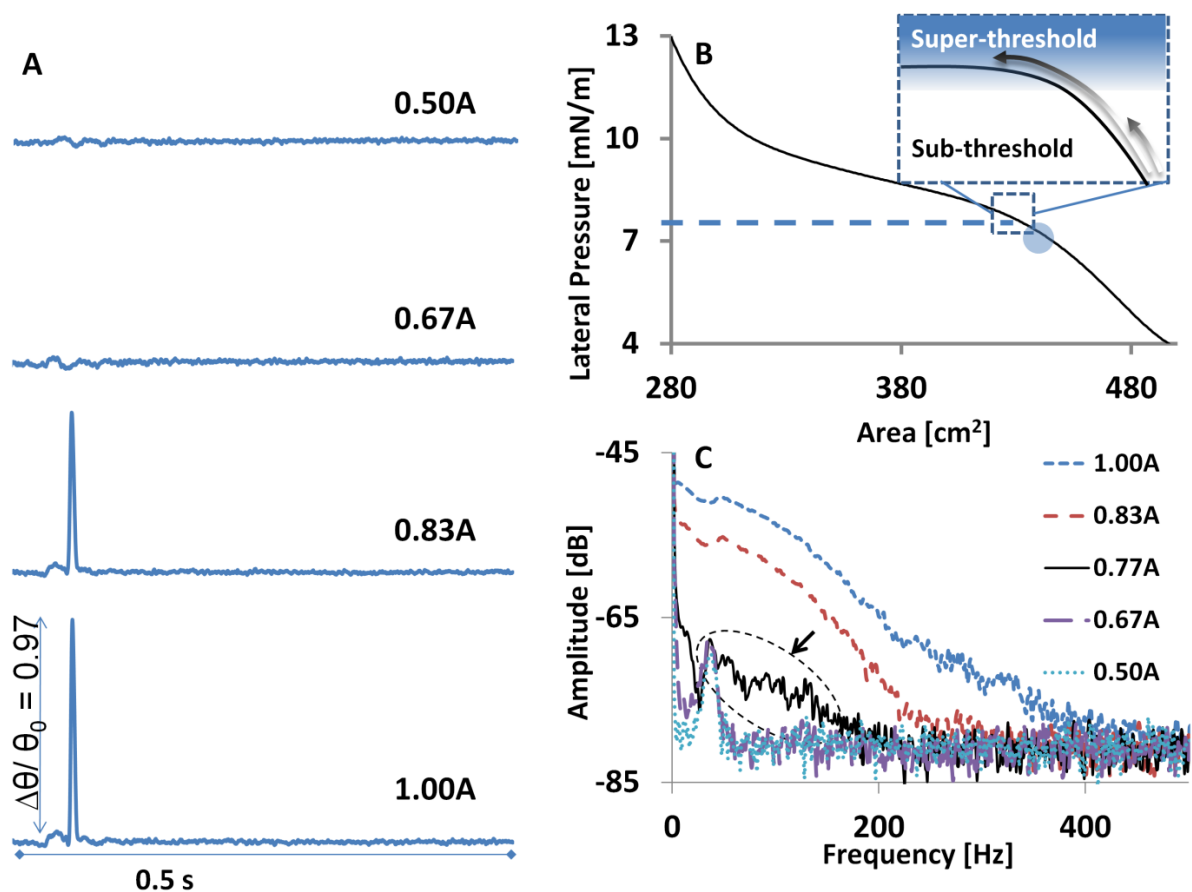

**Figure 3. Threshold for excitation of solitary waves. (A)** The pulses resulting from varying the excitation strength are shown for a fixed state marked by the circle in **(B)**. Clearly a "threshold" for the onset of solitary behavior exists between excitation strength of 0.67 and 0.83A in this state, where maximum displacement amplitude of ~1mm is scaled relatively by 1A. The inset in **(B)** represents the idea of a threshold on a generalized state diagram where solitary pulses can be excited by super-threshold excitation. **(C)** The corresponding average frequency spectrums of the pulses are shown. The arrow points to the onset of nonlinear behavior (see Fig. S4). All the pulses in **(A)** are plotted on same scale for the y-axis representing the variations in FRET parameter. Experiments were performed on a lipid (DPPC) monolayer at a lateral pressure of 7.2mN/m and 21 $^{\circ}$C.

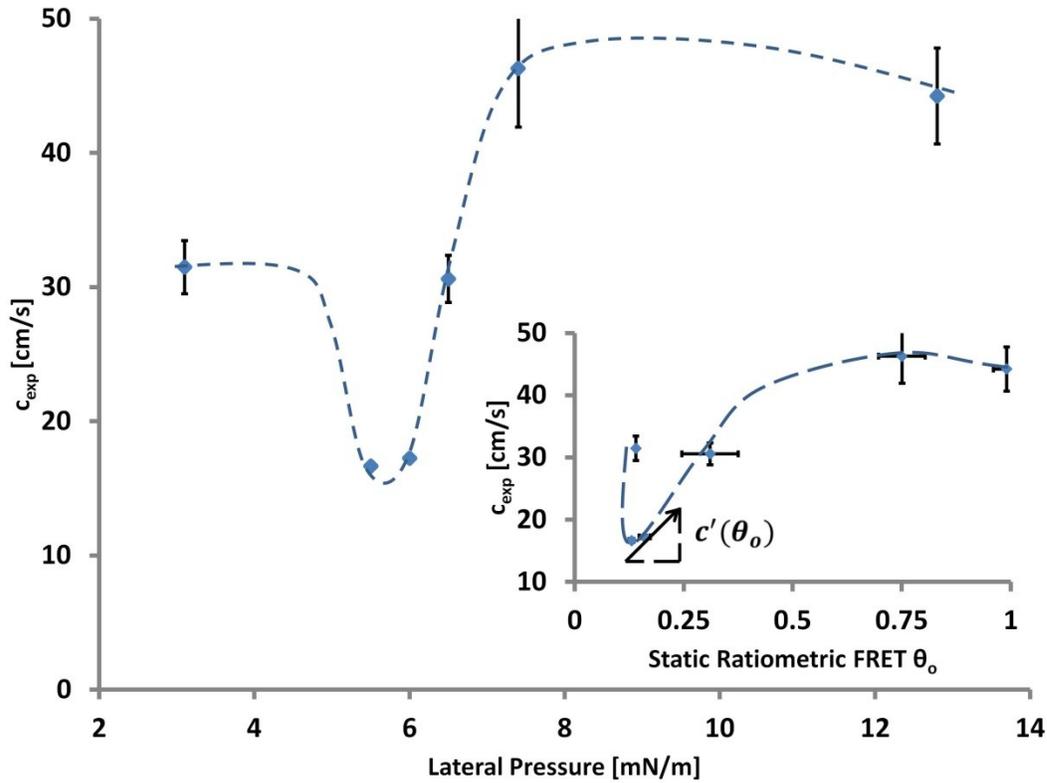

**Figure 4. Non-equilibrium opto-mechanical state diagram.** Mean velocities are plotted as a function of lateral pressure. The minimum in velocity near 6 mN/m corresponds to a maximum in the compressibility of the lipid monolayer. This justifies the interpretation that the pulses measured via FRET are of compressional nature. The inset plots the same data with respect to the equilibrium value of $\theta_o = \left(\frac{I_{535nm}}{I_{605nm}}\right)$ at which the pulse was excited. The slope along any point on this curve is related to the local nonlinearity in the system (see Eq. 3). The experimental velocities were obtained by dividing the distance between excitation and detection (d=1cm) by the time of flight marked by the first deviation in the FRET signal. The pulses were excited at the maximum amplitude of 1A ~ 1mm. The dashed and dotted lines are guides to the eye. Pulses were measured via FRET (see Eq.2 and Fig. S1) in a lipid (DPPC) monolayer at 19 $^o$C and the error bars represent standard deviation over n=10 measurements.

## Supporting Information

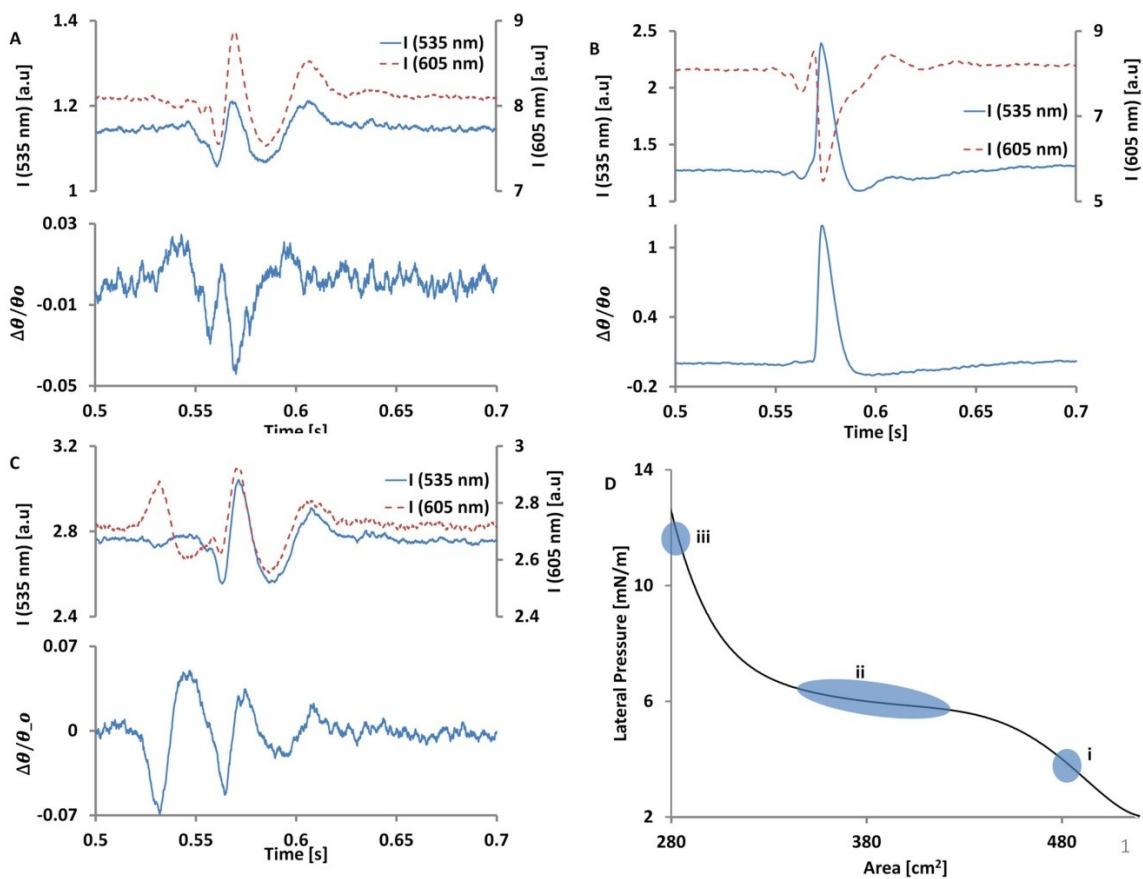

**Figure S1. Optical trace of a propagating pressure pulse.** *(top)* Temporal trace of absolute intensity in the 535 nm and 605 nm channels during a surface pulse at a mean lateral pressure of  (A) 3.1mN/m , (B) 6mN/m and (C) 12.8mN/m marked by region (i), (ii) and (iii) in on the isothermal state diagram in (D) respectively obtained at 19 °C. The FRET variation (plots of $\Delta\theta/\theta_o$) was calculated according to Eq.2 of the main manuscript. The regions where the two signals are out of phase should represent the relative motion between acceptor and donor molecules (Fig. S1) [14,53], while the in-phase variations result from non-FRET processes, such as the vertical motion of the interface with respect to the focus of the microscope. A strong longitudinal mode is clearly visible in (A) even in the absolute intensities *(top)* with complementary rise and fall in the donor and acceptor signal respectively. The longitudinal mode is clearly weaker in (B) and (C). The data presented here is same as in figure 2 of the main manuscript; however note that the scale on y-axis for the pulses varies.

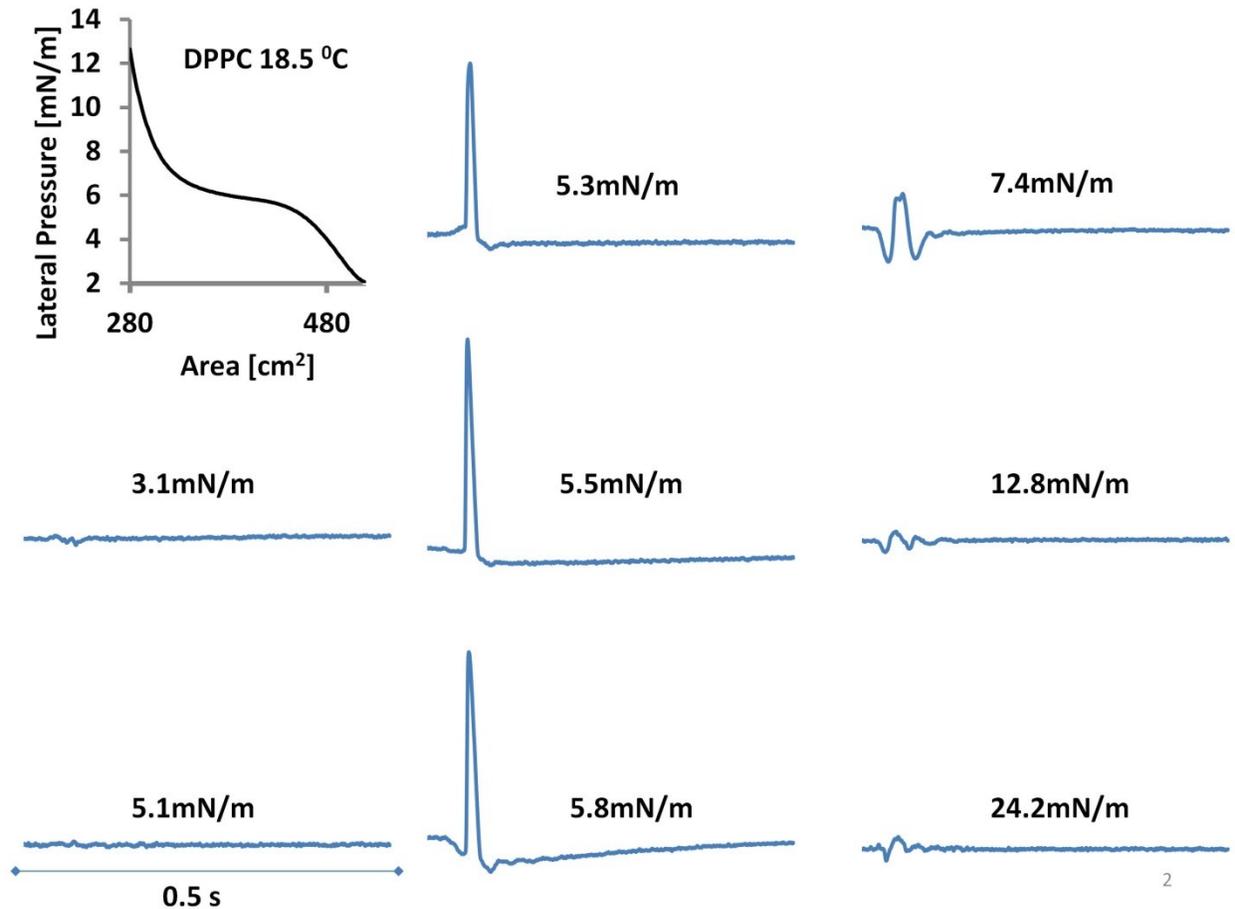

**Figure S2. State dependence of pulse shape**. The effect of mean lateral pressure on the pulse shape is illustrated. The sudden appearance of high amplitude solitary pulses occurs upon changing the pressure from 5.1 to 5.3mN/m. In general, when pulses are excited while proceeding on the state diagram from low to high pressures, it seems characteristic that the first pulses hardly have components below the base line. With increasing pressure, troughs on either side of the sharp crest gradually deepen while the crest splits, blunts and finally disappears completely at high pressures. The pulses were recorded at a distance of 0.7mm from the excitation source at 18.5 $^0$C.

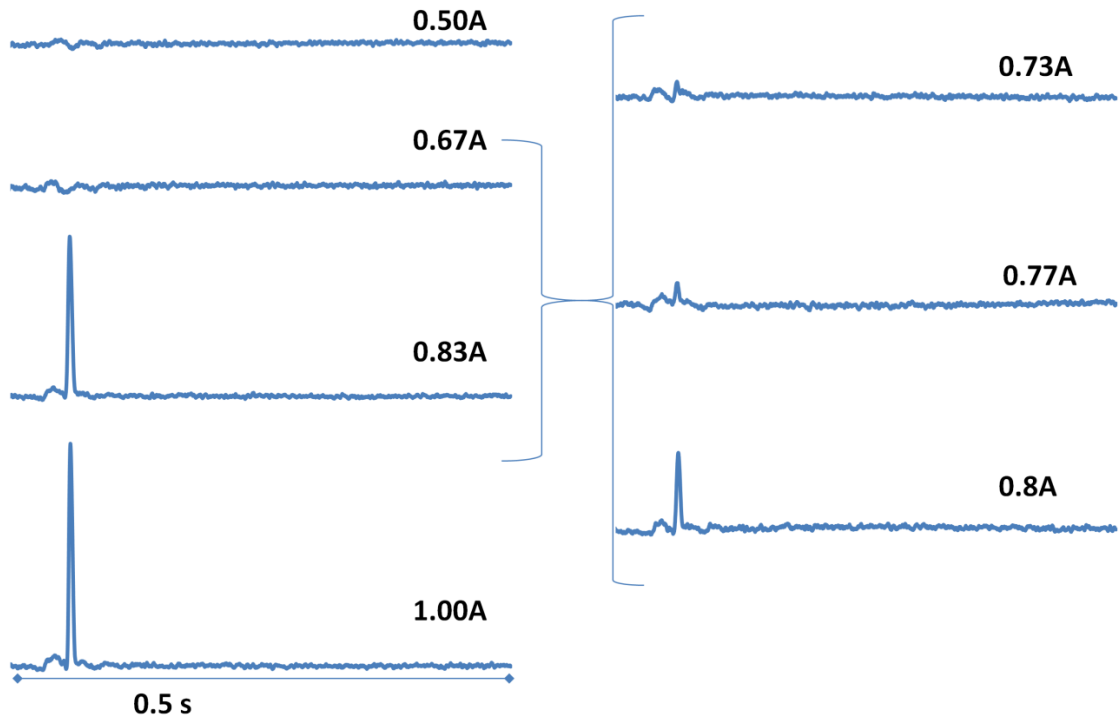

**Figure S3. Resolving threshold excitation strength.** The mechanical excitation strength was gradually increased to identify the exact threshold for the onset of solitary waves. As a result of nonlinear phenomena, one peak of the subthreshold components at 0.73A emerges upon superthreshold excitation. This behavior indicates that the energy is being focused temporally within the pulse. This whole process takes place within a very small range of amplitudes (0.73 to 0.8A) and finally saturates (0.83 to 1A) as explained in the text. The pulses were measured at a mean lateral pressure of 7.2mN/m and 21 °C.

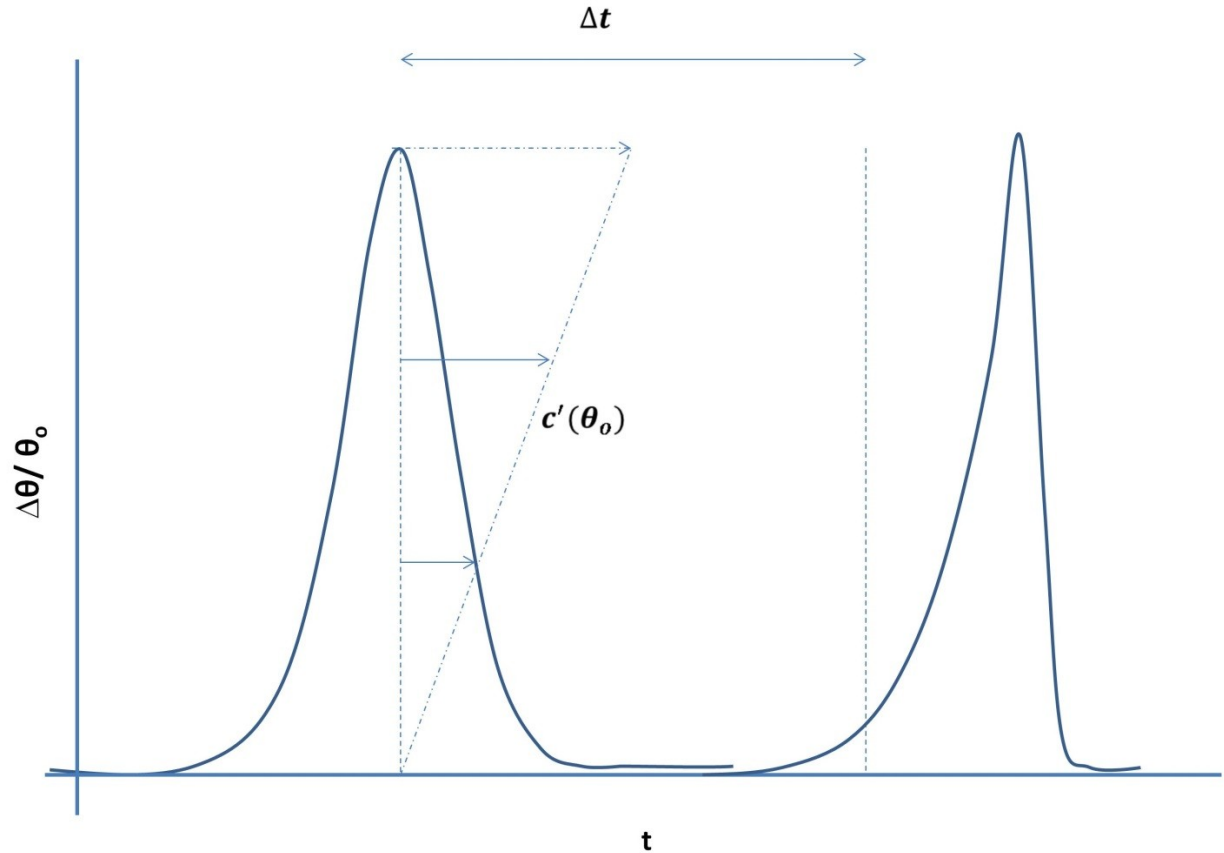

**Figure S4. Nonlinear susceptibility and pulse evolution.** The cartoon depicts the effect of nonlinearity on the pulse shape. In the nonlinear regime *c'* is significant and the velocity depends significantly on the amplitude resulting in different propagation velocities within a single pulse [54]. This causes the pulse shape to evolve, skewing it to a long tail and a steep front, or vice versa, depending on the state-velocity relationship. In terms of frequency spectrum, the steepening is represented by higher frequency components while the tail represents the lower frequencies that account for the resulting broadening of the spectrum. Under "right" conditions dispersion can counterbalance this effect and can thus cause new frequencies to fall back on the "mother" pulse.

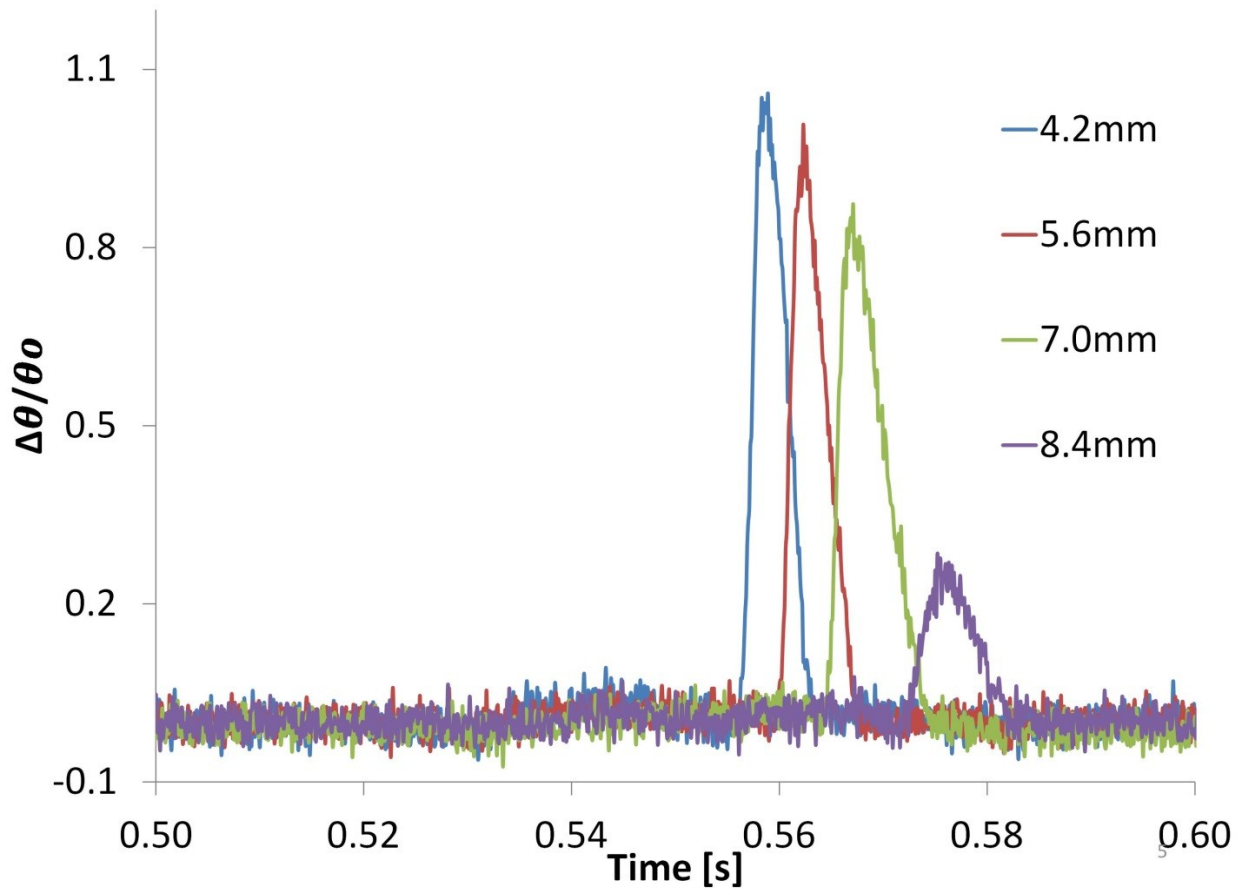

**Figure S5. Decaying Solitary Pulses**. Pulses were measured at different locations by varying the distance (point 'c' in Fig. 1) from the blade. All the experiments shown here were conducted for the same state (7mN/m and 21 °C). On increasing the distance up to 7 mm the amplitude decays moderately. Although at this point we do not know the source of this decay we believe it is driven either by (i) the in-plane spreading (perpendicular to the propagation), a behavior which should be sensitive to lateral confinement of the wave, or (ii) dissipation by viscous friction. Interestingly, further decay of the amplitude ( from 7 mm to 8.4 mm) proceeds nonlinearly. The reason presumably for this behavior is the drop in amplitude to below thresholds level and the system cannot sustain the balance between the dispersive and nonlinear contributions. This once again underlines the solitary nature of these nonlinear pulses as opposed to shock waves that disperse strongly [55]. As evolution of a nonlinear pulse depends on amplitude, dispersion and dissipation, the distance dependence as reported in this figure depends nonlinearly on the exact state of the interface and further studies are underway to systematically analyze these effects.